\documentclass{article}
\usepackage{graphicx}
\usepackage{amsmath}

\input{tcilatex}

\begin{document}

\title{\textbf{Two Time Physics}\\
\textbf{with a Minimum Length}}
\author{W. Chagas-Filho \\
Physics Department, Federal University of Sergipe, Brazil}
\maketitle

\begin{abstract}
We study the possibility of introducing the classical analogue of Snyder's
Lorentz-covariant noncommutative space-time in two-time physics theory. In
the free theory we find that this is possible because there is a broken
local scale invariance of the action. When background gauge fields are
present, they must satisfy certain conditions very similar to the ones first
obtained by Dirac in 1936. These conditions preserve the local and global
invariances of the action and leads to a Snyder space-time with background
gauge fields.
\end{abstract}

\section{Introduction}

Two-Time Physics [1,2,3,4,5,6,7] is an approach that provides a new
perspective for understanding ordinary one-time dynamics from a higher
dimensional, more unified point of view including two time-like dimensions.
This is achieved by introducing a new gauge symmetry that insure unitarity,
causality and absence of ghosts. The new phenomenon in two-time physics is
that the gauge symmetry of the free two-time physics action can be used, by
imposing gauge conditions, to obtain various different actions describing
different free and interacting dynamical systems in the usual one-time
physics, thus uncovering a new layer of unification through higher
dimensions.

An approach to the introduction of background gravitational and gauge fields
in two-time physics was first presented in [7]. In [7], the linear
realization of the $Sp(2,R)$ gauge algebra of two-time physics is required
to be preserved when background gravitational and gauge fields come into
play. To satisfy this requirement, the background gravitational field must
satisfy a homothety condition [7], while in the absence of space-time
gravitational fields the gauge field must satisfy the conditions [7] 
\begin{equation}
X.A(X)=0  \tag{1.1a}
\end{equation}
\begin{equation}
\partial _{M}A^{M}(X)=0  \tag{1.1b}
\end{equation}
\begin{equation}
(X.\partial +1)A_{M}(X)=0  \tag{1.1c}
\end{equation}
which were first proposed by Dirac [8] in 1936. Dirac proposed these
conditions as subsidiary conditions to describe the usual 4-dimensional
Maxwell theory of electromagnetism as a theory in 6 dimensions which
automatically displays $SO(4,2)$ symmetry.

If we recall that in the transition to quantum mechanics $X^{M}\rightarrow
X^{M}$ and $P_{M}\rightarrow i\frac{\partial }{\partial X^{M}}$, we can
rewrite Dirac's conditions (1.1) in the classical form 
\begin{equation}
X.A(X)=0  \tag{1.2a}
\end{equation}
\begin{equation}
P.A(X)=0  \tag{1.2b}
\end{equation}
\begin{equation}
(-iX.P+1)A_{M}(X)=0  \tag{1.2c}
\end{equation}
In this paper we show how a set of subsidiary conditions very similar to
(1.2) can be obtained in the classical Hamiltonian formalism for two-time
physics. As in Dirac's original paper on the $SO(4,2)$-invariant formulation
of electromagnetism [8], the conditions we find in two-time physics are
necessary for the $SO(d,2)$ invariance of the interacting theory. A new
result in this work is that we show that these conditions are also necessary
for a perfect match between the number of physical degrees of freedom
contained in the $(d+2)$-dimensional gauge field and the number of physical
canonical pairs describing the dynamics in the reduced phase space.

The paper is divided as follows. In the next section we review the basic
formalism of two-time physics and show how the $SO(d,2)$ Lorentz generator
for the free 2T action can be obtained from a local scale invariance of the
Hamiltonian. Invariance under this local scale transformation of only the
Hamiltonian reveals that two-time physics can also be consistently
formulated in terms of another set of classical phase space brackets, which
are the classical analogues of the Snyder commutators [9].

In 1947 Snyder proposed a quantized space-time model in a projective
geometry approach to the de Sitter space of momenta with a scale $\theta $
at the Planck scale. In this model, the energy and momentum of a particle
are identified with the inhomogeneous projective coordinates. Then, the
space-time coordinates become noncommutative operators $\hat{x}^{\mu }$
given by the ``translation'' generators of the de Sitter (dS) algebra.
Snyder's space-time has attracted interest in the last few years in
connection with generalizations of special relativity. In particular, it was
pointed out [10] that there is a one-to-one correspondence between the
Snyder space-time and a formulation of de Sitter-invariant special
relativity [11] with two universal invariants, the speed of light $c$ and
the de Sitter radius of curvature $R$.

However, a particle moving in a de Sitter or Anti-de Sitter space-time with
signature $(d-1,1)$, where $d$ is the number of spacelike dimensions, is
only one of the many dual lower-dimensional systems that can be obtained by
imposing gauge conditions on the free two-time physics action (see, for
instance, ref. [4]). Furthermore, a Snyder space-time with signature $%
(d-1,1) $ for a free massless relativistic particle has already been
obtained from the $(d,2)$ space-time of two-time physics by using the Dirac
bracket technique, after imposing gauge conditions to reduce the gauge
invariance of the free 2T action [12]. This shows that in the $(d-1,1)$
space-time there are inertial motions and inertial observers in the Snyder
space-time, giving a principle of relativity for dS/AdS-invariant special
relativity. The results of this work may be used to suggest that the other
universal invariant of dS/AdS special relativity, the radius of curvature $R$%
, can be interpreted as a very large integer multiple of the minimum
spacelike length introduced by the Snyder commutators.

In the treatment of [12], the appearance of Snyder's space-time in the
reduced phase space of the Dirac brackets is a direct consequence of the
fact that the gauge conditions break the conformal $SO(1,2)\sim Sp(2,R)$
gauge invariance of the 2T action, leaving only $\tau $-reparametrization
invariance. Then a length scale induced by the Snyder commutators emerges in
the resulting $(d-1,1)$ space-time, leaving the global scale and conformal
invariances of the gauge-fixed action untouched. To preserve the powerful
unifying properties of 2T physics, and retain the massless particle in a
dS/AdS space-time as one of its gauge-fixed versions, it is then interesting
to investigate the possibility of constructing a Snyder space-time with
signature $(d,2)$, in which the $Sp(2,R)$ gauge invariance and consequently
the full duality properties of the 2T action would be preserved. In this
work we take this task and show that it can be done while also explicitly
preserving the global Lorentz $SO(d,2)$ invariance of the action. These
developments are the content of section two.

In section three we introduce interactions with a background gauge field by
modifying the constraint structure of two-time physics according to the
minimal coupling prescription to electrodynamic gauge fields. We show how a
set of subsidiary conditions very similar to (1.2) emerge after requiring $%
Sp(2,R)$ gauge invariance of the interacting theory and how these conditions
lead to the same Snyder brackets we found in the free theory. Some
concluding remarks appear in section four.

\section{Two-time Physics}

The central idea in two-time physics [1,2,3,4,5,6,7] is to introduce a new
gauge invariance in phase space by gauging the duality of the quantum
commutator $[X_{M},P_{N}]=i\eta _{MN}$. This procedure leads to a symplectic
Sp(2,R) gauge theory. To remove the distinction between position and
momentum we set $X_{1}^{M}=X^{M}$ and $X_{2}^{M}=P^{M}$ and define the
doublet $X_{i}^{M}=(X_{1}^{M},X_{2}^{M})$. The local $Sp(2,R)$ acts as 
\begin{equation}
\delta X_{i}^{M}(\tau )=\epsilon _{ik}\omega ^{kl}(\tau )X_{l}^{M}(\tau ) 
\tag{2.1}
\end{equation}
$\omega ^{ij}(\tau )$ is a symmetric matrix containing three local
parameters and $\epsilon _{ij}$ is the Levi-Civita symbol that serves to
raise or lower indices. The $Sp(2,R)$ gauge field $A^{ij}$ is symmetric in $%
(i,j)$ and transforms as 
\begin{equation}
\delta A^{ij}=\partial _{\tau }\omega ^{ij}+\omega ^{ik}\epsilon
_{kl}A^{lj}+\omega ^{jk}\epsilon _{kl}A^{il}  \tag{2.2}
\end{equation}
The covariant derivative is 
\begin{equation}
D_{\tau }X_{i}^{M}=\partial _{\tau }X_{i}^{M}-\epsilon _{ik}A^{kl}X_{l}^{M} 
\tag{2.3}
\end{equation}
An action invariant under the $Sp(2,R)$ gauge symmetry is 
\begin{equation}
S=\frac{1}{2}\int d\tau (D_{\tau }X_{i}^{M})\epsilon ^{ij}X_{j}^{N}\eta _{MN}
\tag{2.4a}
\end{equation}
After an integration by parts this action can be written as 
\begin{equation*}
S=\int d\tau (\partial _{\tau }X_{1}^{M}X_{2}^{N}-\frac{1}{2}%
A^{ij}X_{i}^{M}X_{j}^{N})\eta _{MN}
\end{equation*}
\begin{equation}
=\int d\tau \lbrack \dot{X}.P-(\frac{1}{2}\lambda _{1}P^{2}+\lambda _{2}X.P+%
\frac{1}{2}\lambda _{3}X^{2})]  \tag{2.4b}
\end{equation}
where $A^{11}=\lambda _{3}$, $A^{12}=A^{21}=\lambda _{2}$, \ $A^{22}=\lambda
_{1}$ and the canonical Hamiltonian is 
\begin{equation}
H=\frac{1}{2}\lambda _{1}P^{2}+\lambda _{2}X.P+\frac{1}{2}\lambda _{3}X^{2} 
\tag{2.5}
\end{equation}
The equations of motion for the $\lambda $'s give the primary [13]
constraints 
\begin{equation}
\phi _{1}=\frac{1}{2}P^{2}\approx 0  \tag{2.6}
\end{equation}
\begin{equation}
\phi _{2}=X.P\approx 0  \tag{2.7}
\end{equation}
\begin{equation}
\phi _{3}=\frac{1}{2}X^{2}\approx 0  \tag{2.8}
\end{equation}
and therefore we can not solve for the $\lambda $'s from their equations of
motion. The values of the $\lambda $'s in action (2.4b) are arbitrary.
Constraints (2.6)-(2.8), as well as evidences of two-time physics, were
independently obtained in [14].

We have introduced the \textbf{weak equality symbol }$\approx $. This is to
emphasize that constraints (2.6)-(2.8) are numerically restricted to be zero
on the submanifold of phase space defined by the constraint equations, but
do not identically vanish throughout phase space [15]. This means, in
particular, that they have nonzero Poisson brackets with the canonical
variables. More generally, two functions $F$ and $G$ that coincide on the
submanifold of phase space defined by constraints $\phi _{i}\approx 0,$ $%
i=1,2,3$ are said to be \textbf{weakly equal }[15] and one writes $F\approx
G $. On the other hand, an equation that holds throughout phase space and
not just on the submanifold $\phi _{i}\approx 0,$ is called \textbf{strong},
and the usual equality symbol is used in that case. It can be demonstrated
[15] that 
\begin{equation}
F\approx G\Longleftrightarrow F-G=c_{i}(X,P)\phi _{i}  \tag{2.9}
\end{equation}

If we consider the Euclidean, or the Minkowski metric as the background
space-time, we find that the surface defined by the constraint equations
(2.6)-(2.8) is trivial. The only metric giving a non-trivial surface,
preserving the unitarity of the theory, and avoiding the ghost problem is
the flat metric with two time-like dimensions [1,2,3,4,5,6,7]. Following
[1,2,3,4,5,6,7] we introduce another space-like dimension and another
time-like dimension and work in a Minkowski space-time with signature $(d,2)$%
.

We use the Poisson brackets 
\begin{equation}
\{P_{M},P_{N}\}=0  \tag{2.10a}
\end{equation}
\begin{equation}
\{X_{M},P_{N}\}=\eta _{MN}  \tag{2.10b}
\end{equation}
\begin{equation}
\{X_{M},X_{N}\}=0  \tag{2.10c}
\end{equation}
where $M,N=0,...,d+1$, and verify that constraints (2.6)-(2.8) obey the
algebra 
\begin{equation}
\{\phi _{1},\phi _{2}\}=-2\phi _{1}  \tag{2.11a}
\end{equation}
\begin{equation}
\{\phi _{1},\phi _{3}\}=-\phi _{2}  \tag{2.11b}
\end{equation}
\begin{equation}
\{\phi _{2},\phi _{3}\}=-2\phi _{3}  \tag{2.11c}
\end{equation}
These equations show that all constraints $\phi $ are first-class [13].
Equations (2.11) represent the symplectic $Sp(2,R)$ gauge algebra of
two-time physics. The 3-parameter local symmetry $Sp(2,R)$ includes $\tau $%
-reparametrizations, generated by constraint $\phi _{1}$, as one of its
local transformations, and therefore the 2T action (2.4) is a generalization
of gravity on the worldline. It corresponds to conformal $SO(2,1)$ gravity
on the worldling [4,14]. Since we have $d+2$ dimensions and $3$ first-class
constraints, only $d+2-3=d-1$ of the canonical pairs $(X_{M},P_{M})$ will
correspond to true physical degrees of freedom.

Action (2.4) also has a global symmetry under Lorentz transformations $%
SO(d,2)$ with generator [1,2,3,4,5,6,7] 
\begin{equation}
L^{MN}=\epsilon ^{ij}X_{i}^{M}X_{j}^{N}=X^{M}P^{N}-X^{N}P^{M}  \tag{2.12}
\end{equation}
It satisfies the space-time algebra 
\begin{equation}
\{L_{MN},L_{RS}\}=\delta _{MR}L_{NS}+\delta _{NS}L_{MR}-\delta
_{MS}L_{NR}-\delta _{NR}L_{MS}  \tag{2.13}
\end{equation}
and is gauge invariant because it has identically vanishing brackets with
the first-class constraints (2.6)-(2.8), $\{L_{MN},\phi _{i}\}=0.$

In one-time physics, a natural way to implement the notion of a minimum
length [16,17,18,19, 20] in theories containing gravity is to formulate
these models on a noncommutative space-time. By a minimum length it is
understood that no experimental device subject to quantum mechanics, gravity
and causality can exclude the quantization of position on distances smaller
than the Planck length [20]. It has been shown [21] that when measurement
processes involve energies of the order of the Planck scale, the fundamental
assumption of locality is no longer a good approximation in theories
containing gravity. The measurements alter the space-time metric in a
fundamental manner governed by the commutation relations $[x_{\mu },p_{\nu
}]=i\eta _{\mu \nu }$ and the classical field equations of gravitation [21].
This in-principle unavoidable change in the space-time metric destroys the
commutativity (and hence locality) of position measurement operators. In the
absence of gravitation locality is restored [21]. This effect of a minimum
length can be modeled by introducing a nonvanishing commutation relation
between the position operators [22].

Let us now consider how the classical analogue of Snyder's noncommutative
space-time can be made to emerge in two-time physics. To arrive at these
classical Snyder brackets we use what can be considered as a broken local
scale invariance of the free 2T action. This local scale invariance is a
symmetry only of the 2T Hamiltonian. It is not a symmetry of the action
because the kinetic term $\dot{X}.P$ in the Legendre transformation, giving
the Lagrangian from the Hamiltonian, is not invariant under this local scale
transformation. This is why we can introduce Snyder brackets in two-time
physics and still preserve the original invariances of the action.

Hamiltonian (2.5) is invariant under the local scale transformations 
\begin{equation}
X^{M}\rightarrow \tilde{X}^{M}=\exp \{\beta \}X^{M}  \tag{2.14a}
\end{equation}
\begin{equation}
P_{M}\rightarrow \tilde{P}_{M}=\exp \{-\beta \}P_{M}  \tag{2.14b}
\end{equation}
\begin{equation}
\lambda _{1}\rightarrow \exp \{2\beta \}\lambda _{1}  \tag{2.14c}
\end{equation}
\begin{equation}
\lambda _{2}\rightarrow \lambda _{2}  \tag{2.14d}
\end{equation}
\begin{equation}
\lambda _{3}\rightarrow \exp \{-2\beta \}\lambda _{3}  \tag{2.14e}
\end{equation}
where $\beta $ is an arbitrary function of $X^{M}(\tau )$ and $P_{M}(\tau )$%
. Keeping only the linear terms in $\beta $ in transformation (2.14), we can
write the brackets 
\begin{equation}
\{\tilde{P}_{M},\tilde{P}_{N}\}=(\beta -1)[\{P_{M},\beta \}P_{N}+\{\beta
,P_{N}\}P_{M}]+\{\beta ,\beta \}P_{M}P_{N}  \tag{2.15a}
\end{equation}
\begin{equation*}
\{\tilde{X}_{M},\tilde{P}_{N}\}=(1+\beta )[\eta _{MN}(1-\beta
)-\{X_{M},\beta \}P_{N}]
\end{equation*}
\begin{equation}
+(1-\beta )X_{M}\{\beta ,P_{N}\}-X_{M}X_{N}\{\beta ,\beta \}  \tag{2.15b}
\end{equation}
\begin{equation}
\{\tilde{X}_{M},\tilde{X}_{N}\}=(1+\beta )[X_{M}\{\beta
,X_{N}\}-X_{N}\{\beta ,X_{M}\}]+X_{M}X_{N}\{\beta ,\beta \}  \tag{2.15c}
\end{equation}
for the transformed canonical variables. If we choose $\beta =\phi _{1}=%
\frac{1}{2}P^{2}$ $\approx 0$ in equations (2.15) and compute the brackets
on the right side using the Poisson brackets (2.10), we find the expressions 
\begin{equation}
\{\tilde{P}_{M},\tilde{P}_{N}\}=0  \tag{2.16a}
\end{equation}
\begin{equation}
\{\tilde{X}_{M},\tilde{P}_{N}\}=(1+\frac{1}{2}P^{2})[\eta _{MN}(1-\frac{1}{2}%
P^{2})-P_{M}P_{N}]  \tag{2.16b}
\end{equation}
\begin{equation}
\{\tilde{X}_{M},\tilde{X}_{N}\}=-(1+\frac{1}{2}P^{2})(X_{M}P_{N}-X_{N}P_{M})
\tag{2.16c}
\end{equation}
We see from the above equations that, on the constraint surface defined by
constraints (2.6)-(2.8), brackets (2.16) reduce to 
\begin{equation}
\{\tilde{P}_{M},\tilde{P}_{N}\}=0  \tag{2.17a}
\end{equation}
\begin{equation}
\{\tilde{X}_{M},\tilde{P}_{N}\}=\eta _{MN}-P_{M}P_{N}  \tag{2.17b}
\end{equation}
\begin{equation}
\{\tilde{X}_{M},\tilde{X}_{N}\}=-(X_{M}P_{N}-X_{N}P_{M})  \tag{2.17c}
\end{equation}
To impose $\phi _{1}=\frac{1}{2}P^{2}\approx 0$ strongly at the end of the
computation of brackets (2.16), the expressions for the corresponding Dirac
brackets should be used in place of the Poisson brackets. However, for the
special case $\beta =\phi _{1}=\frac{1}{2}P^{2}\approx 0$ we can use the
property\ [15] of the Dirac brackets that, on the first-class constraint
surface, 
\begin{equation}
\{G,F\}_{D}\approx \{G,F\}  \tag{2.18}
\end{equation}
when $G$ is a first-class constraint and $F$ is an arbitrary function of the
canonical variables. This justifies the use of Poisson brackets to arrive at
(2.17).

Now, keeping the same order of approximation used to arrive at brackets
(2.15), that is, retaining only the linear terms in $\beta $, transformation
equations (2.14a) and (2.14b) read 
\begin{equation}
\tilde{X}^{M}=\exp \{\beta \}X^{M}=(1+\beta )X^{M}  \tag{2.19a}
\end{equation}
\begin{equation}
\tilde{P}_{M}=\exp \{-\beta \}P_{M}=(1-\beta )P_{M}  \tag{2.19b}
\end{equation}
Using again the same function $\beta =\phi _{1}=\frac{1}{2}P^{2}\approx 0$
in equations (2.19), we write them as 
\begin{equation}
\tilde{X}^{M}=X^{M}+\frac{1}{2}P^{2}X^{M}  \tag{2.20a}
\end{equation}
\begin{equation}
\tilde{P}_{M}=P_{M}-\frac{1}{2}P^{2}P_{M}  \tag{2.20b}
\end{equation}
or, equivalently, 
\begin{equation}
\tilde{X}^{M}-X^{M}=C_{i}^{M}(X,P)\phi _{i}  \tag{2.21a}
\end{equation}
\begin{equation}
\tilde{P}_{M}-P_{M}=D_{M}^{i}(X,P)\phi _{i}  \tag{2.21b}
\end{equation}
with $C_{1}^{M}=X^{M},$ $C_{2}^{M}=C_{3}^{M}=0$ and $D_{M}^{1}=-P_{M},$ $%
D_{M}^{2}=D_{M}^{3}=0$. Equations (2.21) are obviously in the form (2.9) and
so we can write 
\begin{equation}
\tilde{X}^{M}\approx X^{M}  \tag{2.22a}
\end{equation}
\begin{equation}
\tilde{P}_{M}\approx P_{M}  \tag{2.22b}
\end{equation}
Using these weak equalities in brackets (2.17) we rewrite them as

\begin{equation}
\{P_{M},P_{N}\}\approx 0  \tag{2.23a}
\end{equation}
\begin{equation}
\{X_{M},P_{N}\}\approx \eta _{MN}-P_{M}P_{N}  \tag{2.23b}
\end{equation}
\begin{equation}
\{X_{M},X_{N}\}\approx -(X_{M}P_{N}-X_{N}P_{M})  \tag{2.23c}
\end{equation}
to emphasize that these brackets are valid only on the constraint surface
defined by constraints (2.6)-(2.8). But, as we saw above, the non-trivial
surfaces corresponding to constraints (2.6)-(2.8) require a space-time with
signature $(d,2)$.

Brackets (2.23) are the classical 2T equivalent of the Lorentz-covariant
Snyder commutators [9], which were proposed in 1947 as a way to solve the
ultraviolet divergence problem in quantum field theory by introducing a
minimum space-time length. In the canonical quantization procedure, where
brackets are replaced by commutators according to the rule 
\begin{equation*}
\lbrack commutator]=i\{bracket\}
\end{equation*}
the 2T brackets (2.23) will lead directly to a Lorentz-covariant
noncommutative space-time for two-time physics, thus implementing the notion
of a minimum length in the $(d+2)$-dimensional space-time for this theory.

The Snyder brackets (2.23) give an equally valid description of two-time
physics at the classical level. If we compute the bracket $\{L_{MN},L_{RS}\}$
using the Snyder brackets we find that the same space-time algebra (2.13) is
reproduced. This implies that the Snyder brackets (2.23) preserve the global 
$SO(d,2)$ Lorentz invariance of action (2.4). Since $SO(d,2)$ contains scale
as well as conformal transformations we see that, although we may introduce
a scale at the Planck length using the Snyder brackets (2.23), global scale
and conformal invariances still exist. This is because to arrive at the
Snyder brackets (2.23) we have used the local scale invariance (2.14) of the
2T Hamiltonian, which is a broken scale invariance from the Lagrangian point
of view.

If we compute the brackets $\{L_{MN},\phi _{i}\}$ using (2.23) to verify the
gauge invariance of $L_{MN}$ in a phase space with Snyder brackets, we find
that the $\{L_{MN},\phi _{i}\}$ identically vanish, proving that $L_{MN}$ is
gauge invariant in this phase space. Computing the algebra of constraints
(2.6)-(2.8) using (2.23) we arrive at the expressions 
\begin{equation}
\{\phi _{1},\phi _{2}\}=-2\phi _{1}+4\phi _{1}^{2}  \tag{2.24a}
\end{equation}
\begin{equation}
\{\phi _{1},\phi _{3}\}=-\phi _{2}+2\phi _{1}\phi _{2}  \tag{2.24b}
\end{equation}
\begin{equation}
\{\phi _{2},\phi _{3}\}=-2\phi _{3}+\phi _{2}^{2}  \tag{2.24c}
\end{equation}
which show that the first-class property of constraints (2.6)-(2.8) is
preserved by brackets (2.23). Equations (2.24) are the realization of the $%
Sp(2,R)$ gauge algebra of two-time physics in a phase space with Snyder
brackets. Equations (2.24) exactly reproduce the gauge algebra (2.11) if we
take the linear approximation on the right side..

Notice that $L_{MN}$ explicitly appears with a minus sign in the right hand
side of the Snyder bracket (2.23c), establishing a connection between the
global $SO(d,2)$ Lorentz invariance of action (2.4) and the local scale
invariance (2.14) of Hamiltonian (2.5).

The new result obtained in this section is that the classical and free
two-time physics theory can also be consistently formulated in a phase space
where the Snyder brackets (2.23) are valid. In the next section we will see
that this remains true in the presence of a background gauge field $A_{M}(X)$
when a set of subsidiary conditions very similar to (1.2) are satisfied.

\section{2T Physics with Gauge Fields}

To introduce a background gauge field $A_{M}(X)$ we modify the free action
(2.4b) according to the usual minimal coupling prescription to gauge fields, 
$P_{M}\rightarrow P_{M}-A_{M}$. The interacting 2T action in this case is
then 
\begin{equation}
S=\int d\tau \{\dot{X}.P-[\frac{1}{2}\lambda _{1}(P-A)^{2}+\lambda
_{2}X.(P-A)+\frac{1}{2}\lambda _{3}X^{2}]\}  \tag{3.1}
\end{equation}
where the Hamiltonian is 
\begin{equation}
H=\frac{1}{2}\lambda _{1}(P-A)^{2}+\lambda _{2}X.(P-A)+\frac{1}{2}\lambda
_{3}X^{2}  \tag{3.2}
\end{equation}
The equations of motion for the multipliers now give the constraints 
\begin{equation}
\phi _{1}=\frac{1}{2}(P-A)^{2}\approx 0  \tag{3.3}
\end{equation}
\begin{equation}
\phi _{2}=X.(P-A)\approx 0  \tag{3.4}
\end{equation}
\begin{equation}
\phi _{3}=\frac{1}{2}X^{2}\approx 0  \tag{3.5}
\end{equation}
The Poisson brackets between the canonical variables and the gauge field are 
\begin{equation}
\{X_{M},A_{N}\}=0  \tag{3.6a}
\end{equation}
\begin{equation}
\{P_{M},A_{N}\}=-\frac{\partial A_{N}}{\partial X_{M}}  \tag{3.6b}
\end{equation}
\begin{equation}
\{A_{M},A_{N}\}=0  \tag{3.6c}
\end{equation}

Computing the algebra of constraints (3.3)-(3.5) using the Poisson brackets
(2.9) and (3.6) we obtain the equations 
\begin{equation*}
\{\phi _{1},\phi _{2}\}=-2\phi _{1}+(P^{M}-A^{M})\frac{\partial }{\partial
X^{M}}(X.A)-(P-A).A
\end{equation*}
\begin{equation}
-X^{M}\frac{\partial }{\partial X^{M}}[(P-A).A]-X^{M}\frac{\partial }{%
\partial X^{M}}(\frac{1}{2}A^{2})  \tag{3.7a}
\end{equation}
\begin{equation}
\{\phi _{1},\phi _{3}\}=-\phi _{2}  \tag{3.7b}
\end{equation}
\begin{equation}
\{\phi _{2},\phi _{3}\}=-2\phi _{3}  \tag{3.7c}
\end{equation}
Equations (3.7)\ exactly reproduce the $Sp(2,R)$ gauge algebra (2.11) when
the conditions 
\begin{equation}
X.A=0  \tag{3.8a}
\end{equation}
\begin{equation}
(P-A).A=0  \tag{3.8b}
\end{equation}
\begin{equation}
\frac{1}{2}A^{2}=0  \tag{3.8c}
\end{equation}
hold. Condition (3.8a) is the first of Dirac's subsidiary conditions (1.2)
on the gauge field. Conditions (3.8b) and (3.8c) are, however, different
from (1.2b) and (1.2c). When conditions (3.8) hold, the $Sp(2,R)$ gauge
algebra (2.11) is reproduced by constraints (3.3)-(3.5). Thus, when (3.8)
holds, the only possible space-time metric associated with constraints
(3.3)-(3.5) giving a non-trivial surface and avoiding the ghost problem is
the flat metric with two time-like dimensions.

Now that we have seen that action (3.1) has a local $Sp(2,R)$ gauge
invariance when conditions (3.8) hold, we may consider the question of which
is the $SO(d,2)$ Lorentz generator for action (3.1). A possible answer is
obtained if we use the minimal coupling prescription $P_{M}\rightarrow
P_{M}-A_{M}$ in the expression for $L_{MN}$ in the free theory, thus
obtaining in the interacting theory 
\begin{equation*}
L_{MN}^{I}=X_{M}(P_{N}-A_{N})-X_{N}(P_{M}-A_{M})
\end{equation*}
\begin{equation*}
=(X_{M}P_{N}-X_{N}P_{M})-(X_{M}A_{N}-X_{N}A_{M})
\end{equation*}
\begin{equation}
=L_{MN}-S_{MN}  \tag{3.9}
\end{equation}
We can even use the first-class gauge function $\beta =\phi _{1}=\frac{1}{2}%
(P-A)^{2}\approx 0$ and formally construct a set of Snyder brackets for the
interacting theory in which $L_{MN}^{I}$ appears on the right side of the
bracket $\{X_{M},X_{N}\}$, exactly in the same way as $L_{MN}$ appears on
the right side of (2.23c) in the free theory. However, it can be verified
(using Poisson brackets) that the global transformations generated by $%
S_{MN}=X_{M}A_{N}-X_{N}A_{M}$ identically vanish, 
\begin{equation}
\delta X_{R}=\frac{1}{2}\epsilon _{MN}\{S_{MN},X_{R}\}=0  \tag{3.10a}
\end{equation}
\begin{equation}
\delta P_{R}=\frac{1}{2}\epsilon _{MN}\{S_{MN},P_{R}\}=0  \tag{3.10b}
\end{equation}
\begin{equation}
\delta A_{R}=\frac{1}{2}\epsilon _{MN}\{S_{MN},A_{R}\}=0  \tag{3.10c}
\end{equation}
The Lorentz generator for the interacting theory is then effectively
identical to $L_{MN}$ in the free theory. This agrees with Dirac's
interpretation of the conformal $SO(4,2)$ symmetry of Maxwell's theory as
being the Lorentz symmetry in 6 dimensions. This was also pointed out, but
in a rather unclear way, in reference [7] (see section four of [7]).

The above conclusion implies that $L_{MN}$ must be invariant under the gauge
transformations generated by constraints (3.3)-(3.5). Using the Poisson
brackets (2.9) and (3.6) we find the equations 
\begin{equation*}
\{L_{MN},\phi _{1}\}=X_{M}\frac{\partial }{\partial X_{N}}[(P-A).A]+X_{M}%
\frac{\partial }{\partial X_{N}}(\frac{1}{2}A^{2})
\end{equation*}
\begin{equation*}
-X_{N}\frac{\partial }{\partial X_{M}}[(P-A).A]-X_{N}\frac{\partial }{%
\partial X_{M}}(\frac{1}{2}A^{2})+P_{M}\frac{\partial }{\partial P_{N}}[%
(P-A).A]
\end{equation*}
\begin{equation}
-P_{N}\frac{\partial }{\partial P_{M}}[(P-A).A]  \tag{3.11a}
\end{equation}
\begin{equation}
\{L_{MN},\phi _{2}\}=X_{M}\frac{\partial }{\partial X_{N}}(X.A)-X_{N}\frac{%
\partial }{\partial X_{M}}(X.A)  \tag{3.11b}
\end{equation}
\begin{equation}
\{L_{MN},\phi _{3}\}=0  \tag{3.11c}
\end{equation}
We see from the above equations that $L_{MN}$ is gauge invariant, $%
\{L_{MN},\phi _{i}\}=0$, when conditions (3.8) are valid. Action (3.1),
complemented with the subsidiary conditions (3.8), gives therefore a
consistent classical Hamiltonian description of two-time physics with
background gauge fields in a phase space with the Poisson brackets (2.9) and
(3.6). But, as we saw in section two, there is another Hamiltonian
description of two-time physics based in a phase space with the Snyder
brackets (2.23). Let us then consider this Hamiltonian formulation in the
case when background gauge fields are present.

Since the Lorentz generator $L_{MN}$ in the interacting theory is identical
to the one in the free theory, the form (2.23) of the Snyder brackets must
also be preserved in the interacting theory because $L_{MN}$ explicitly
appears in the right side of (2.23c). This creates a mathematical difficulty
because the gauge function $\beta =\frac{1}{2}P^{2}$ we used to arrive at
(2.23) in the free theory is no longer a first-class function on the
constraint surface defined by (3.3)-(3.5). Consequently, equations (2.9) and
(2.18) can not be used. To solve this difficulty we incorporate conditions
(3.8) as new constraints for the interacting theory.

Combining conditions (3.8) with constraints (3.3)-(3.5), we get the
irreducible [15] set of constraints 
\begin{equation}
\phi _{1}=\frac{1}{2}P^{2}\approx 0  \tag{3.12}
\end{equation}
\begin{equation}
\phi _{2}=X.P\approx 0  \tag{3.13}
\end{equation}
\begin{equation}
\phi _{3}=\frac{1}{2}X^{2}\approx 0  \tag{3.14}
\end{equation}
\begin{equation}
\phi _{4}=X.A\approx 0  \tag{3.15}
\end{equation}
\begin{equation}
\phi _{5}=P.A\approx 0  \tag{3.16}
\end{equation}
\begin{equation}
\phi _{6}=\frac{1}{2}A^{2}\approx 0  \tag{3.17}
\end{equation}
Note that Dirac's equations (1.2a) and (1.2b) are now reproduced by
constraints $\phi _{4}$ and $\phi _{5}$ above. But now there is a clear
meaning for the third condition: the gauge field must remain massless.
Constraints (3.12)-(3.17) obey the $Sp(2,R)$ gauge algebra (2.11) together
with the equations 
\begin{equation}
\{\phi _{1},\phi _{4}\}=-P_{M}\frac{\partial }{\partial X_{M}}(X.A)\approx 0
\tag{3.18a}
\end{equation}
\begin{equation}
\{\phi _{1},\phi _{5}\}=-P_{M}\frac{\partial }{\partial X_{M}}(P.A)\approx 0
\tag{3.18b}
\end{equation}
\begin{equation}
\{\phi _{1},\phi _{6}\}=-P_{M}\frac{\partial }{\partial X_{M}}(\frac{1}{2}%
A^{2})\approx 0  \tag{3.18c}
\end{equation}
\begin{equation}
\{\phi _{2},\phi _{4}\}=-X_{M}\frac{\partial }{\partial X_{M}}(X.A)\approx 0
\tag{3.18d}
\end{equation}
\begin{equation}
\{\phi _{2},\phi _{5}\}=P.A-X_{M}\frac{\partial }{\partial X_{M}}%
(P.A)\approx 0  \tag{3.18e}
\end{equation}
\begin{equation}
\{\phi _{2},\phi _{6}\}=-X_{M}\frac{\partial }{\partial X_{M}}(\frac{1}{2}%
A^{2})\approx 0  \tag{3.18f}
\end{equation}
\begin{equation}
\{\phi _{3},\phi _{4}\}=0  \tag{3.18g}
\end{equation}
\begin{equation}
\{\phi _{3},\phi _{5}\}=X.A\approx 0  \tag{3.18h}
\end{equation}
\begin{equation}
\{\phi _{3},\phi _{6}\}=0  \tag{3.18i}
\end{equation}
Equations (2.11) together with equations (3.18) show that all constraints
(3.12)-(3.17) are first-class constraints.

Hamiltonian (3.2) will be invariant under the local scale transformations
(2.14) when the gauge field effectively transforms as 
\begin{equation}
A_{M}\rightarrow \tilde{A}_{M}=\exp \{-\beta \}A_{M}  \tag{3.19}
\end{equation}
Using this local scale invariance we can again construct the same brackets
(2.15). On the constraint surface defined by equations (3.12)-(3.17) we can
use again equation (2.9) and the property (2.18) of the Dirac bracket and
choose $\beta =\phi _{1}=\frac{1}{2}P^{2}\approx 0$ to arrive, by the same
steps described in the previous section, at the same Snyder brackets (2.23).

Finally, let us consider the role of conditions (3.8). As in the free
theory, the first-class constraints (3.12)-(3.14) reduce the number of
physical canonical pairs $(X,P)$ to be $d-1$. We introduced a gauge field $%
A_{M}(X)$ with $d+2$ components, but as a consequence of (3.8) now there are 
$3$ first-class constraints (3.15)-(3.17) acting on these $d+2$ components.
These constraints can be used to reduce the number of independent components
of the gauge field to be $d+2-3=d-1$, creating a perfect match of the number
of independent components of the gauge field with the number of physical
canonical pairs. It is this perfect match that preserves the local $Sp(2,R)$
invariance in the presence of the gauge field.

\section{Concluding remarks}

In this investigation we considered the possibility of introducing a minimum
length in the classical two-time physics theory by constructing its
Hamiltonian formulation in a phase space with Snyder brackets. It makes
sense to try to introduce this minimum length in two time physics because
the action is a generalization of gravity on the world-line and gravity
introduces additional uncertainties in the quantum position measurement
process.

We saw that it is possible to introduce a minimum length in the free theory
and in the presence of background gauge fields, while at the same time
preserving the usual symmetries of two-time physics, because of the
existence of a broken local scale invariance which is a symmetry only of the
Hamiltonian. We clarified a previous observation of the fact that the global 
$SO(d,2)$ Lorentz generator in the presence of background gauge fields is
identical to the one in the free theory and exposed the connection of this
Lorentz generator with the concept of a minimum length We also revealed the
mechanism for the preservation of the local $Sp(2,R)$ invariance of the
action, which consists in a perfect match, in the Hamiltonian formalism,
between the number of physical canonical pairs describing the dynamics and
the number of physical components in the gauge field.

\noindent

\noindent

\end{document}